\documentclass[aps,prb,twocolumn,groupedaddress]{revtex4-1}
\usepackage{graphics}
\usepackage{amsmath}
\usepackage{float}
\usepackage{times}
\begin{document}

\title{Classical nucleation theory of ice nucleation: second-order correction of \\thermodynamic parameters}
\author{Chaohong Wang}
\author{Jianyang Wu}
\author{Hao Wang}
\thanks{}
\email{Corresponding author: h\_wang@xmu.edu.cn}
\author{Zhisen Zhang}
\thanks{}
\email{Corresponding author: zhangzs@xmu.edu.cn}
\affiliation{Department of Physics, Research Institute for Biomimetics and Soft Matter and Fujian Provincial Key Laboratory for Soft Functional Materials Research, Xiamen University, Xiamen 361005, PR China}
\begin{abstract}
Accurate estimate of nucleation rate is crucial for the study of ice nucleation and ice-promoting/anti-freeze strategies. 
Within the framework of Classical Nucleation Theory (CNT), the estimate of ice nucleation rate 
is very sensitive to thermodynamic parameters, such as chemical potential difference between water and ice $\Delta \mu$ and ice-water interfacial free energy $\gamma$. 
However, even today, there are still many contradictions and approximations in the estimating of 
these thermodynamic parameters, introducing large uncertainty to the estimate of the ice nucleation rate. Herein, starting from the basic concepts, for a general solid-liquid crystallization system, 
we expand the Gibbs-Thomson (GT) equation to second order, and derive the second-order analytical formulas of 
$\Delta \mu$, $\gamma$ and nucleation barrier $\Delta G$ with combining molecular dynamics (MD) simulations. These formulas describe well the temperature dependence of these thermodynamic parameters. 
Our results can provide a method of estimating $\Delta \mu$, $\gamma$ and $\Delta G$.
\end{abstract}
\maketitle

\section{INTRODUCTION}
Water freezing is a ubiquitous phenomenon in nature, with important consequences in a variety of environments, 
including climate, transportation infrastructure, biological cell, and industrial production. 
There has been a lot of important researches on ice formation \cite{moore2011structural,lupi2017role,matsumoto2002molecular,li2013ice}. 
However, there is much debate about the mechanism of water freezing. Water freezing is a phase change process that crystallizes 
from supercooled water, which is affected by many and even uncertain factors \cite{RN8}. The nucleation of ice 
is a key step through the process. The homogeneous nucleation can be well described by the CNT \cite{RN182,RN181,kelton1991crystal}. 
According to the CNT, ice embryos are formed by thermal fluctuations in supercooled liquid water. When the size of 
ice embryos exceeds the critical size, ice nuclei will spontaneously grow. In this process, 
ice embryos are required to overcome the nucleation barrier $\Delta G$, which represents the resistance 
to nucleation. For general crystallization systems, in the case of a spherical solid nucleus forming from the supercooled liquid, the nucleation 
barrier can be expressed as
\begin{align}
    \Delta G=\dfrac{16 \pi \gamma^{3}}{3\left(\rho_{s} \Delta \mu\right)^{2}}
\end{align}
where $\gamma$ is the solid-liquid interfacial free energy, 
$\Delta \mu$ is the chemical potential difference between liquid and solid, and 
$\rho_{s}$ is the particle number density of the solid nuclei. Once nucleation barrier 
and kinetic parameters are known, the homogeneous nucleation rate can be estimated, 
which can describe the probability of homogeneous ice nucleation under a set of ambient conditions. 
The homogeneous nucleation rate $R_{hom}$ varies with the nucleation temperature $T$ following the Arrhenius 
equation \cite{beenakker1991solid}
\begin{align}
    R_{h o m}=A_{h o m} \cdot \exp \left(-\dfrac{\Delta G}{k_{B} T}\right)
\end{align}
where $A_{hom}$ is the kinetic prefactor, 
$k_{B}$ is Boltzmann constant. $\Delta G$ is in the exponent term, which greatly 
affects the value of $R_{hom}$. Moreover, according to Eq.(1), the nucleation 
barrier is sensitive to the thermodynamic parameters, like $\gamma$ and $\Delta \mu$. 
Therefore, obtaining accurate thermodynamic parameters is important for 
estimating the nucleation rate.

However, at present, there are barely reliable experimental methods to directly measure such micro 
thermodynamic parameters. Currently, the ice-water interfacial free energy can be estimated 
by using MD simulations or based on the fitting of CNT to measured nucleation rates \cite{RN7}. 
However, due to the ambiguous concept and various estimate methods, there is a large variation 
in the reported estimates of $\gamma$ that span between $25$ and $35\ mJ/m^{2}$ at melting temperature \cite{RN186}. 
And through research, it is found that $\gamma$ is strongly dependent on 
temperature, and there are many different reports on this dependence \cite{RN107}. 
In reports that estimate $\gamma$ as a linear parameter, literature estimates of $d\gamma/dT$ vary between $0.10$ and $0.25\ mJ/(m^{2} \cdot K)$ \cite{granasy2002interfacial,Chukin2010HomogeneousIN}. 
Since the temperature dependence of $\gamma$ is difficult to accurately estimate, 
$\gamma$ is usually approximated as a constant which measured at melting temperature in many cases. 
For the chemical potential difference between water and ice, so far, 
people often use an approximate formula to describe its relationship with supercooling. 
There are also accurate methods to estimate $\Delta \mu$, like thermodynamic integration \cite{frenkel2001understanding}. 
However, the mathematical relationship with supercooling is still vague. Consequently, fuzzy thermodynamic parameters 
and improper mathematical approximations bring great uncertainty to the estimate of nucleation barrier and nucleation rate.

In this paper, to avoid the above-mentioned problem of unclear quantitative relationship and approximate treatment, 
we use the thermodynamic methods to expand the GT equation, and taking advantages of MD simulations, theoretically give the analytical 
formulas of $\Delta \mu$, $\gamma$ and $\Delta G$.

\section{THEORY}
Considering that a solid cluster is in equilibrium with its supercooled liquid phase at temperature $T_{c}$, 
this also means that the temperature $T_{c}$ is the melting point of the cluster. 
For the mechanical equilibrium, the curved interface exerts a pressure difference to the cluster. 
It can be described by Laplace’s equation
\begin{align}
    p_{s}-p_{l}=\gamma K
\end{align}
where $p_{s}$ and $p_{l}$ are the pressure of the solid phase and liquid phase, 
respectively, $\gamma$ is the solid-liquid interfacial free energy, and $K$ is the curvature of the interface. For the chemical equilibrium, 
these two phases have the same chemical potential
\begin{align}
    \mu_{s}\left(p_{s}, T_{c}\right)=\mu_{l}\left(p_{l}, T_{c}\right)
\end{align}
where $\mu_{s}$ and $\mu_{l}$ are the chemical potential of the solid phase and liquid phase, 
respectively. According to Gibbs-Duhem (GD) relation: $d \mu=-SdT+vdp$, where $S$ is the molecular entropy and $v$ is the molecular volume, 
for an incompressible phase, in general, the chemical potential of solid phase at the pressure $p_{s}$ can be expressed by using the pressure $p_{l}$ as a reference
\begin{align}
    \mu_{s}\left(p_{s}, T_{c}\right)=\mu_{s}\left(p_{l}, T_{c}\right)+v_{s} \gamma K
\end{align}
where $v_{s}$ is the solid molecular volume, which is also the reciprocal of $\rho_{s}$. 
Applying Eq.(4) and Eq.(5), we obtain
\begin{align}
    \mu_{l}\left(p_{l}, T_{c}\right)-\mu_{s}\left(p_{l}, T_{c}\right)=v_{s} \gamma K
\end{align}
The above equation is the chemical potential difference between liquid phase and 
solid phase in mother phase environment, namely, $\Delta \mu$. 
For both phases, we integrate the GD relation from current condition to coexistence condition.

For liquid phase, at different temperatures, the pressure changes very little, with regarding $p_{l}$ as a constant, 
we obtain
\begin{align}
    \mu_{l}\left(p_{l}, T_{m}\right)-\mu_{l}\left(p_{l}, T_{c}\right) = \int_{T_{c}}^{T_{m}}-S_{l}\left(T\right) d T
\end{align}
For solid phase, since the additional interface pressure, the pressure change is not negligible
\begin{align}
    \mu_{s}\left(p_{l}, T_{m}\right)-\mu_{s}\left(p_{s}, T_{c}\right) = \int_{T_{c}}^{T_{m}}-S_{s}\left(T\right) d T+\int_{p_{s}}^{p_{l}} v_{s} d p
\end{align}
At the melting temperature $T_{m}$, two phases’ chemical potential are equal
\begin{align}
    \mu_{s}\left(p_{l}, T_{m}\right)=\mu_{l}\left(p_{l}, T_{m}\right)
\end{align}
Now substituting the expressions Eq.(3), Eq.(4) and Eq.(9) 
into Eq.(8)$-$Eq(7). Simplifying equation with $\Delta S\left(T\right)=S_{l}\left(T\right)-S_{s}\left(T\right)=S_{l}\left(T_{m}\right)-S_{s}\left(T_{m}\right)=\Delta H_{m}/T_{m}$, where $\Delta H_{m}$ is the melting enthalpy of solid phase, 
then writing
\begin{align}
    \Delta T = T_{m}-T_{c} = \frac{v_{s}\gamma K T_{m}}{\Delta H_{m}}
\end{align}
This is the GT equation, which describes the melting point depression of the solid cluster. Combining Eq.(6), it can be rewritten as
\begin{align}
    \Delta \mu = v_{s} \gamma K = \frac{\Delta H_{m}}{T_{m}} \Delta T
\end{align}
The Eq.(11) is often used to estimate $\Delta \mu$.

However, it is not very appropriate to treat the difference in entropy of the liquid phase and solid phase as a constant 
because the temperature dependencies of the entropy of the liquid phase and the solid phase are not consistent, especially with the increase of supercooling, this difference will gradually deviate from $S_{l}\left(T_{m}\right)-S_{s}\left(T_{m}\right)$.
Therefore, in order to more accurately describe the behavior of entropy with temperature, 
in Eq.(7) and Eq.(8), Taylor expand entropy at $T_{m}$ to linear term, we obtain
\begin{align}
    \mu_{l}\left(p_{l}, T_{m}\right)-\mu_{l}\left(p_{l}, T_{c}\right) = \int_{T_{c}}^{T_{m}}-S_{l}\left(T\right) d T \notag \\ = -\int_{T_{c}}^{T_{m}}\left[S_{l}\left(T_{m}\right)+\left(\frac{\partial S_{l}}{\partial T}\right)_{p}\left(T-T_{m}\right)\right] d T
\end{align}
\begin{align}
    \mu_{s}\left(p_{l}, T_{m}\right)-\mu_{s}\left(p_{s}, T_{c}\right) = \int_{T_{c}}^{T_{m}}-S_{s}\left(T\right) d T+\int_{p_{s}}^{p_{l}} v_{s} d p \notag \\ = -\int_{T_{c}}^{T_{m}}\left[S_{s}\left(T_{m}\right)+\left(\frac{\partial S_{s}}{\partial T}\right)_{p}\left(T-T_{m}\right)\right] d T +\int_{p_{s}}^{p_{l}} v_{s} d p
\end{align}
Similarly, simplifying Eq.(13)$-$Eq(12) with $C_{p}=T(\partial S/\partial T)_{p}$, then writing
\begin{align}
    \Delta \mu=v_{s} \gamma K=\dfrac{\Delta H_{m} \Delta T-\dfrac{\Delta C}{2}(\Delta T)^{2}}{T_{m}}
\end{align}
where $\Delta C$ is difference of constant pressure heat capacity between liquid and 
solid ($C_{p}^{l}-C_{p}^{s}$) at $T_{m}$ and $p_{l}$. Comparing with Eq.(11), 
this is a second-order expansion of GT equation. It is worth mentioning that, from the derivation process, 
this formula is also applicable to other incompressible solid-liquid systems. 
Although there is also second-order GT equation \cite{RN46}, which regards $\gamma$ as a constant and expands $\Delta T$ into a polynomial of $K$. 
However, in this paper, we regard $\gamma$ as a variable. The second-order GT equation shows the relationship between interfacial free energy $\gamma$, 
supercooling $\Delta T$ and interface curvature $K$. Therefore, for estimating $\Delta \mu$ and $\gamma$, 
we need to get the value of $\Delta C$ and the relationship between the curvature (for spherical cluster $K=2/r$, $r$ is equilibrium radius) of the interface 
and the melting point of solid cluster. In the following content, we apply this formula to ice nucleation system through MD simulation to estimate $\Delta \mu$ and $\gamma$.

\section{SIMULATION DETAILS}
To obtain the value of $\Delta C$, we need to get the constant pressure heat capacity 
of water and ice at the temperature of $T_{m}$ and the pressure of 1 bar respectively. 
The constant pressure heat capacity $C_{p}$ is defined as
\begin{align}
    C_{p} & = \left(\dfrac{\partial H}{\partial T}\right)_{p}
\end{align}
where $H$ is the enthalpy of the bulk phase system. Thus, it is necessary to calculate the enthalpy at different temperatures and then take the derivative to get the isobaric heat capacity at the melting temperature. 

We use TIP4P/ice model \cite{RN45} to build the cuboid system of ice and water, each of them contains 4800 water molecules. 
TIP4P/ice was designed to reproduce the melting temperature, the densities, 
and the coexistence curves of several ice phases. Some of its properties are as follows in TABLE I. 
We set a series of temperature (255, 260, 265, 270, 275, 280 and 285 K) around melting temperature, 
and perform NPT GROMACS \cite{van2005gromacs} MD simulations for each temperature. Long-range electrostatic interaction is 
calculated by using the smooth Particle Mesh Ewald method \cite{RN187} and the van der Waals interaction is modeled 
using a Lennard-Jones potential. Both the LJ and the real part of the Coulombic interactions truncated 
at 1.3 \AA. The rigid geometry of the water model and periodic boundary conditions are preserved. All simulations are run at the constant 
pressure of 1 bar, using an isotropic Parrinello-Rahman barostat \cite{RN188} and at constant temperature using 
the velocity-rescaling thermostat \cite{RN190}. The MD time-step was set to 2 fs and each system was equilibrated about 0.2 ns at 200 K. 
All MD simulations run for 40 nanoseconds and the last 20 ns of each simulation is taken as a statistical sample.
\begin{table}[H] 
\caption{Some properties of TIP4P/ice model. $T_{m}$ is melting temperature; $\rho_{Ih}$, density of Ih ice; $\Delta H_{m}$, the melting enthalpy.}
\begin{ruledtabular}
\begin{tabular}{cccc}
Model&$T_{m}(K)$&$\rho_{Ih}(g/cm^{3})$&$\Delta H_{m}(kJ/mol)$\\\hline
TIP4P/ice&270 \cite{RN154}&0.906 \cite{RN45}&5.40 \cite{RN45}\\
\end{tabular}
\end{ruledtabular}
\end{table}
    
\section{RESULTS AND DISCUSSION}
\subsection{Isobaric heat capacity}
The mean of enthalpy in each system is shown in Fig.\ 1. 
The linear change of enthalpy to temperature indicates that there is no significant temperature 
dependence of isobaric heat capacity in this temperature range. 
Derived from the data, $C_{p}^{water}=97.5\ J/(mol \cdot K)$ and $C_{p}^{ice}=58.0\ J/(mol \cdot K)$. Therefore, 
$\Delta C$ equals $39.5\ J/(mol \cdot K)$, which is close to $41.8\ J/(mol \cdot K)$ from the calculation for TIP4P/2005 \cite{RN148} and the experimental value $40.1\ J/(mol \cdot K)$ \cite{RN75}.
\begin{figure}[H]
\includegraphics{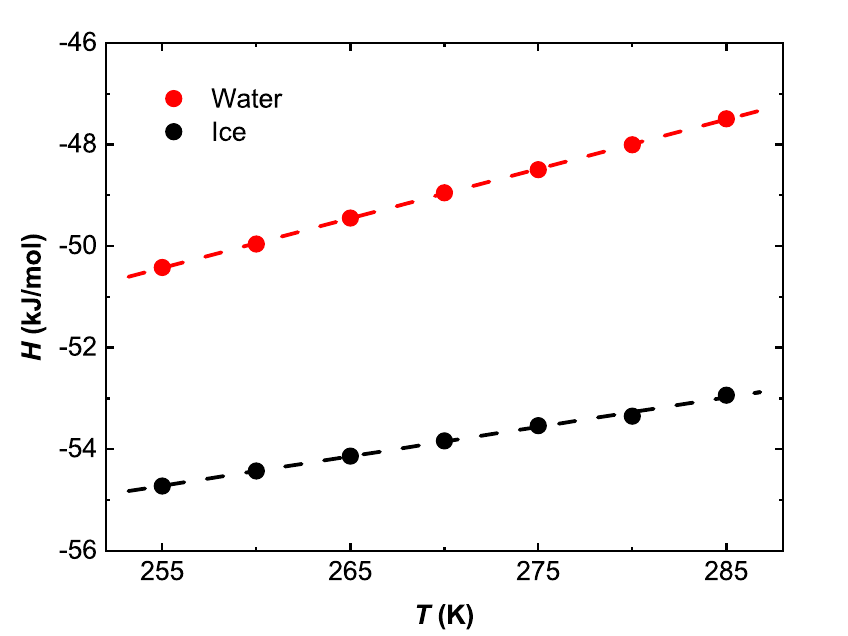}
\caption{Enthalpy of ice and water at various temperatures. The dash lines are linear fitted lines.}
\end{figure}
\subsection{Chemical potential difference between water and ice, $\Delta \mu$}
Comparing Eq.(11) with Eq.(14), both of them can describe the temperature dependence of $\Delta \mu$, and their difference is that the latter has an additional second-order correction term. 
Whereby the value of $\Delta C$, we compare the two approaches in Fig.\ 2. 
Under low subcooling, they are not much different. 
However, with increase of supercooling, the correction term can not be ignored. And the conclusion that the second-order value is 
smaller than the first-order approximation is consistent with the results obtained by thermodynamic integration \cite{RN109,RN117}.
\begin{figure}[H]  
\includegraphics{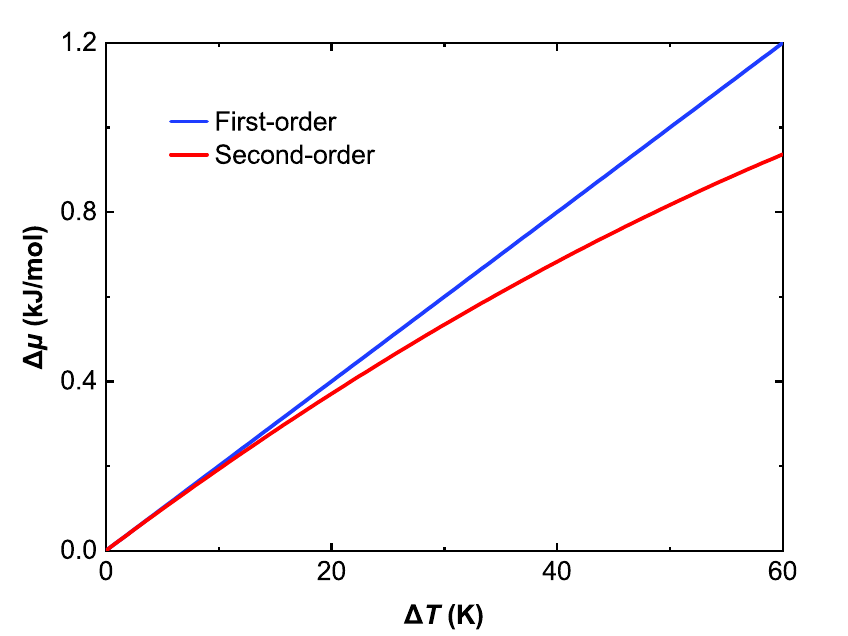}
\caption{Two descriptions of the temperature dependence on the $\Delta \mu$. The blue line and red line denote first-order $\Delta \mu$ and second-order $\Delta \mu$, respectively.}
\end{figure}
\subsection{Melting point and equilibrium radius}
For a general solid-liquid crystallization system, the size dependence of melting point of solid cluster has been studied a lot. 
There is a relationship that the melting point depression $\Delta T$ of a nanoparticle varies inversely to its equilibrium radius $r$ (critical radius) in many results \cite{RN121,RN86,RN90}, 
which can be explained by GT equation (Eq.(10)). In recent work \cite{RN168}, this relationship reappeared in ice-water system as shown in Fig.\ 3, 
and it is confirmed in a recent experimental work \cite{bai2019probing}. 
We also use the seeding technique \cite{RN121} to verify this relationship (see the Supplemental Material).
\begin{figure}[H]
\includegraphics{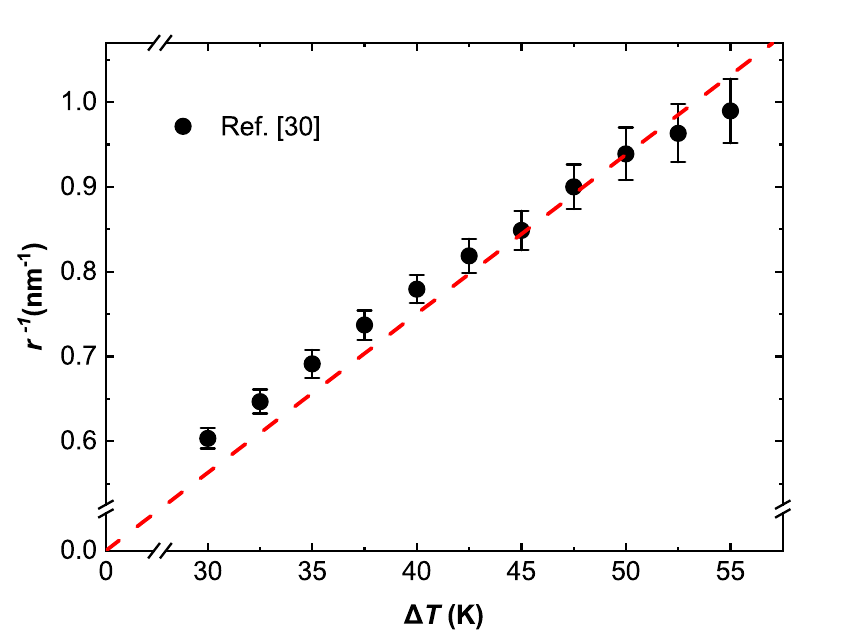}
\caption{The relationship between the inverse of radius of the equilibrium ice cluster and $\Delta T$. The data represented by the black dot is processed from Niu and Parrinello's data \cite{RN168}, $r$ is estimated by critical nuclei sizes $N_{c}$ and cluster density $\rho$ in the paper. The red dash line is the fitted line fitted in inverse proportion.}
\end{figure}
Whereas, in Eq.(10), $\Delta T$ and $r$ are not strictly inversely proportional, since $\gamma$ is a variable that 
changes with $\Delta T$ or $r$. And compared to the second-order GT equation, Eq.(10) neglects the error caused by high supercooling. 
Thus, combining with these two factors and Fig.\ 3, we make a reasonable presumption to make an equation describes the inverse proportional relationship by rewriting the GT equation as 
\begin{align}
    \Delta T = \frac{2 \gamma_{0} T_{m} v_{s}}{r \Delta H_{m}}
\end{align}
where $\gamma_{0}$ is a constant with the same dimension as $\gamma$, and its physical meaning is discussed in the next subsection. 
\subsection{Interfacial free energy, $\gamma$}
Now, with the knowledge of the second-order GT equation and the relationship between $\Delta T$ and $r$, 
substituting the expression Eq.(16) into Eq.(14) and eliminate $r$
\begin{align}
    \gamma = \gamma_{0}-\frac{\gamma_{0} \Delta C}{2 \Delta H_{m}}\left(T_{m}-T\right)
\end{align} 
According to the equation, interfacial free energy is a linear function of temperature. 
This is qualitative agreement with experimental and MD simulations estimates of the behavior of $\gamma$ with $T$ \cite{RN107}. When $T=T_{m}$, we get $\gamma=\gamma_{0}$. So the physical meaning of 
$\gamma_{0}$ is the interfacial free energy at the melting temperature. Its value depends on the slope of the fitted line in the Fig.\ 3. 
We get $\gamma_{0}=26.8\ mJ/m^{2}$ through Eq.(16). It is within the aforementioned normal range of ice-water interfacial free energy. 
In addition, the slope of $\gamma$ is $0.10\ mJ/(m^{2} \cdot K)$ which is in the aforementioned range of $d\gamma/dT$. 
Similarly, substituting the expression Eq.(16) into Eq.(14) and eliminate $\Delta T$
\begin{align}
    \frac{\gamma}{\gamma_{0}} = 1-\frac{\gamma_{0} \Delta C T_{m} v_{s}}{r\left(\Delta H_{m}\right)^{2}} = 1-\frac{\delta}{r}
\end{align}
where $\delta$ is a constant and equals $1.9$ \AA. Noticeably, this formula is similar to the Tolman’s equation \cite{RN144}, 
and $\delta$ is on the same order of magnitude as Tolman length in previous work \cite{RN184}. However, Eq.(18) shows the relationship between 
the interfacial free energy and the critical radius under different temperatures, not the curvature 
correction at a specific temperature.

Different from the interfacial free energy on a certain crystal plane, all the interfacial free energy discussed above are average interfacial free energy of spherical ice-water interface. 
This concept is consistent with the interfacial free energy in CNT.
\subsection{Nucleation barrier, $\Delta G$}
The relationship between the nucleation barrier and the supercooling has always been concerned by 
the researchers who study nucleation. While this relationship is often estimated based on approximate thermodynamic parameters, 
which regards $\Delta \mu \propto \Delta T$ and $\gamma$ as a constant. Inserting them to Eq.(1) leads to
\begin{align}
    \Delta G = \frac{16 \pi T_{m}^{2} v_{s}^{2} \gamma_{0}^{3}}{3\left(\Delta H_{m}\right)^{2}(\Delta T)^{2}}
\end{align}
This formula is widely used as the basis for estimating nucleation barrier and nucleation rate. 
Now the second-order corrections for $\Delta \mu$ and $\gamma$ are obtained. Inserting Eq.(14), Eq.(16) and Eq.(17) into Eq.(1) leads to
\begin{align}
    \Delta G = \frac{16 \pi T_{m}^{2} v_{s}^{2} \gamma_{0}^{3}}{3\left(\Delta H_{m}\right)^{2}(\Delta T)^{2}}-\frac{8 \pi T_{m}^{2} v_{s}^{2} \Delta C \gamma_{0}^{3}}{3\left(\Delta H_{m}\right)^{3} \Delta T}
\end{align}
Compared with the first-order nucleation barrier (Eq.19), this is second-order nucleation barrier. The mathematical form of the new formula is $a/(\Delta T)^{2}-b/\Delta T$, rather than $a/(\Delta T)^2$. 
To verify this new formula, we use these two formulas to estimate the values of $\Delta G$ using the parameters $\gamma_{0}=26.8\ mJ/m^{2}$ and $\Delta C=39.5\ J/(mol \cdot K)$ and compare them with Niu and Parrinello's data as shown in Fig.\ 4. From the figure, 
it can be seen that the new formula estimate better (see TABLE II for details). We also directly use these two formulas to fit the data to compare the fitting effect in the Supplemental Material. 
Due to the existence of the correction term, the second-order nucleation barrier is smaller than the first-order. 
This is important for accurate nucleation barrier predictions since a small difference in the nucleation barrier can lead 
to several orders of magnitude difference in the nucleation rate. With increase of supercooling, the absolute value of the correction term decreases, but its proportion increases. 
Therefore, whether under high or low supercooling, the effect of the correction term can not be neglected. 
\begin{figure}
\includegraphics{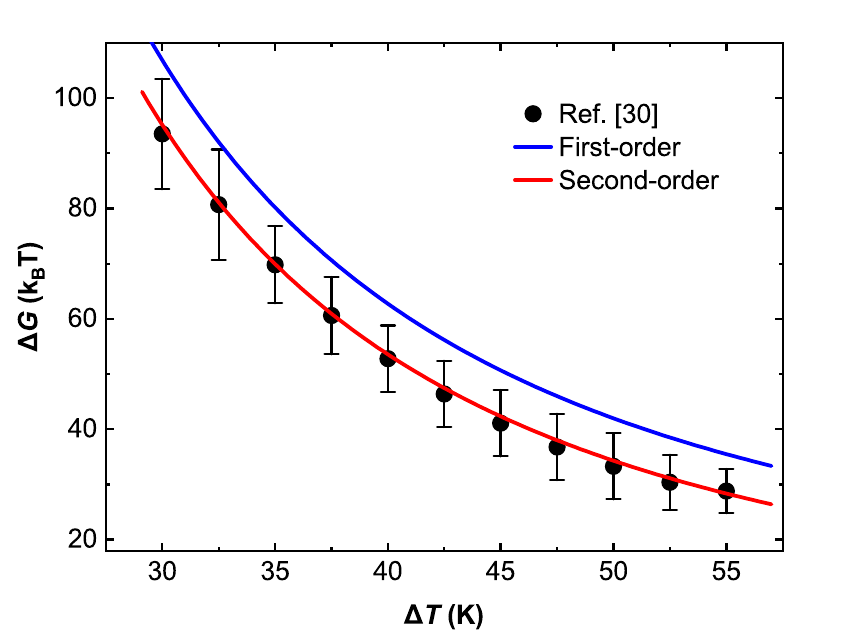}
\caption{Temperature dependence of $\Delta G$. The data represented by the black dot is from Niu and Parrinello's data \cite{RN168}. The blue line and red line denote first-order nucleation barrier and second-order nucleation barrier respectively.}
\end{figure}
\begin{table}
\caption{Details of estimate. Using thermodynamic parameters to estimate the nucleation barrier, where $\Delta G_{1}$ and $\Delta G_{2}$ are estimated by first-order nucleation barrier and second-order nucleation barrier respectively, $\Delta \Delta G_{1}$ is the difference between $\Delta G_{1}$ and $\Delta G$, $\Delta \Delta G_{2}$ is the same, and $\beta$ is $1/(k_{B}T)$.}
\begin{ruledtabular}
\begin{tabular}{ccccccc}
$\Delta T(K)$&30&35&40&45&50&55\\
$\beta \Delta G$ \ Ref. \cite{RN168}      &93.5  &69.8  &52.8  &41.1  &33.3  &28.8\\\hline
$\beta \Delta G_{1}$                      &106.9 &80.2  &62.7  &50.7  &42.0  &35.5\\
$\beta \Delta \Delta G_{1}$               &13.4  &10.4  &9.9   &9.6   &8.7   &6.7 \\\hline
$\beta \Delta G_{2}$	                  &95.2  &69.9  &53.6  &42.3  &34.3  &28.4\\
$\beta \Delta \Delta G_{2}$               &1.7   &0.1   &0.8   &1.2   &1.0   &-0.4\\
\end{tabular}
\end{ruledtabular}
\end{table}

\section{SUMMARIZE}
In this paper, we demonstrated that using a second-order GT equation corrects the temperature dependence of $\Delta \mu$ and $\gamma$ in ice nucleation system. 
For $\Delta \mu$, the second-order expression is more accurate than commonly used approximate expression. 
Compared with fitting CNT to estimate $\gamma$, the second-order expression explains the linear dependence of $\gamma$ on temperature more clearly. 
Normally these formulas are also applicable to other incompressible crystallization systems. 
Furthermore, combined with CNT, the second-order $\Delta G$ is obtained, which takes into account the more realistic temperature-dependent behavior of $\Delta \mu$ and $\gamma$ and 
makes accurately predicting the nucleation barrier and nucleation rate possible. 
Our work provides theoretical guidance for studying ice nucleation and other nucleation systems.

\section{ACKNOWLEDGEMENTS}
This work is financially supported by the National Natural Science Foundation of China (Grant Nos. 11904300, 11772278, 11502221 and 51907171), 
the Jiangxi Provincial Outstanding Young Talents Program (Grant No. 20192BCBL23029), the Fundamental Research Funds for the Central Universities (Xiamen University: Grant Nos. 20720180014 and 20720180018). 
Y. Yu and Z. Xu from Information and Network Center of Xiamen University for the help with the high-performance computer.

\bibliography{manuscript}

\end{document}